\documentclass[reprint,prb,aps,superscriptaddress,showpacs]{revtex4-1}

\usepackage{amsmath}
\usepackage{epsfig}
\usepackage{graphicx}
\usepackage{float}
\usepackage{blindtext}
\usepackage{color}
\usepackage[nameinlink]{cleveref}
\usepackage[colorlinks=true,allcolors=blue,citecolor=blue]{hyperref}
\usepackage{natbib}
\usepackage{smartref}
\usepackage[utf8]{inputenc}

\begin{document}

\title{$d_{xz/yz}$ Orbital Subband Structures and Chiral Orbital Angular Momentum in the (001) Surface States of SrTiO$_3$}

\author{Shoresh Soltani}
\affiliation{Institute of Physics and Applied Physics, Yonsei University, Seoul 120-749, Republic of Korea}
\affiliation{Center for Correlated Electron Systems, Institute for Basic Science (IBS), Seoul 08826, Republic of Korea}

\author{Soohyun Cho}
\affiliation{Institute of Physics and Applied Physics, Yonsei University, Seoul 120-749, Republic of Korea}
\affiliation{Center for Correlated Electron Systems, Institute for Basic Science (IBS), Seoul 08826, Republic of Korea}

\author{Hanyoung Ryu}
\affiliation{Center for Correlated Electron Systems, Institute for Basic Science (IBS), Seoul 08826, Republic of Korea}
\affiliation{Department of Physics and Astronomy, Seoul National University (SNU), Seoul 08826, Republic of Korea}

\author{Garam Han}
\affiliation{Center for Correlated Electron Systems, Institute for Basic Science (IBS), Seoul 08826, Republic of Korea}
\affiliation{Department of Physics and Astronomy, Seoul National University (SNU), Seoul 08826, Republic of Korea}

\author{Beomyoung Kim}
\affiliation{Advanced Light Source, Lawrence Berkeley National Laboratory, Berkeley, CA 94720, USA}

\author{Dongjoon Song}
\affiliation{Electronic and Photonics Research Institute, National Institute of Advanced Industrial Science and Technology (AIST), Tsukuba 305-8568, Japan}

\author{Timur K. Kim}
\affiliation{Diamond Light Source, Harwell Campus, Didcot OX11 0DE, United Kingdom}
\author{Moritz Hoesch}
\affiliation{Diamond Light Source, Harwell Campus, Didcot OX11 0DE, United Kingdom}

\author{Changyoung Kim}
\email{changyoung@snu.ac.kr}
\affiliation{Center for Correlated Electron Systems, Institute for Basic Science (IBS), Seoul 08826, Republic of Korea}
\affiliation{Department of Physics and Astronomy, Seoul National University (SNU), Seoul 08826, Republic of Korea}

\begin{abstract}
We have performed angle resolved photoemission spectroscopy (ARPES) experiments on the surface states of SrTiO$_3$(001) using linearly and circularly polarized light to investigate the subband structures of out-of-plane $d_{xz/yz}$ orbitals and chiral orbital angular momentum (OAM). The data taken in the first Brillouin zone reveal new subbands for $d_{xz/yz}$ orbitals with Fermi wave vectors of 0.25 and 0.45 $\mathrm{\AA}^{-1}$ in addition to the  previously reported ones. As a result, there are at least two subbands for all the Ti 3d t$_{2g}$ orbitals. Our circular dichroism ARPES data is suggestive of a chiral OAM structure in the surface states and may provide clues to the origin of the linear Rashba-like surface band splitting.
\\
\\
PACS number(s): 79.60.-i, 79.60.Bm, 73.20.At, 73.21.Fg
\end{abstract}

\date{\today}
\maketitle

\section{Introduction}
Studying two dimensional electron gases (2DEGs) in the surfaces and interfaces of transitional metal oxides has been an interesting topic during the past decades due to the intriguing properties of confined electronic states. In particular, the creation and control of a 2DEG on surfaces of SrTiO$_3$ (STO) and in STO/Al$_2$O$_3$ interfaces has ignited  extensive research \cite{Ohtomo,Meevasana}. The confined electronic states show unique and interesting properties including superconductivity, magnetism, multiferroicity and enhanced Seebeck coefficients~\cite{Herranz,Moetakef,Lin,Banerjee,Ohta}. Such novel phenomena make STO important in oxide electronics applications. To have deeper understanding of these phenomena and fabricate practical devices, we also need realistic models and experimental observations of confined states within a narrow surface potential. All these make studies of metallic states on STO important in the field of solid state physics.

Several theoretical and experimental studies of the surface states of STO have been performed so far~\cite{Mattheiss,Ben Shalom,Caviglia,Santander-Syro1,King,Walker1,Rodel,Plumb,Walker2,Khalsa,Kim,Popovic,Caviglia,Berner,Chang,Li Li,Chen,Wang,Chang2}. Especially, exploiting its surface sensitivity, ARPES has been used to directly measure the band structures of STO metallic states. For instance, subband structures and their orbital characters have been investigated~\cite{Santander-Syro1}. Other than ARPES studies, it is known from transport measurements that each subband has a small Rashba-like splitting with linear and non-linear terms.~\cite{Nakamura}

While the band dispersion and the origin of 2DEG at the surface seem to be well studied, there are still issues to be resolved. For example, the origin of the surface band splitting deduced from the transport experiments is not settled~\cite{Nakamura}. Different approaches have been used to explain it within the standard Rashba or unconventional Rashba models~\cite{Zhong,Khalsa,Kim,Panjin Kim,King,Santander-Syro2,Walker3}. Zhong \emph{et. al.} suggested that the spin orbit coupling (SOC) effect at the crossing point of the $d_{xy}$ and $d_{yz}$ (or $d_{zx}$) bands can result in a Rashba spin splitting with a cubic term.~\cite{Zhong} However, the Rashba effect of $t_{2g}$ bands was phenomenologically treated which, for example, cannot explain the complex spin or orbital angular momentum structures~\cite{King}. On the other hand, Kim \emph{et. al.}~\cite{Panjin Kim} used an approach based on the orbital Rashba effect model~\cite{J.H. Park,S.R Park} and claimed that the approach not only explains the linear and cubic momentum terms but also predicts the chiral spin and orbital angular momentum. In addition, a recent spin-ARPES study~\cite{Santander-Syro2} shows the existence of a giant spin splitting (100 meV) and suggested a magnetic order on the surface within a non-Rashba picture while another spin-ARPES measurement~\cite{Walker3} rejects the existence of such spin splitting and treats it as an unconventional Rashba splitting. These raise new questions on the origin of the band splitting and possibility for time-reversal symmetry breaking. As for the band structure, previous data show only one elliptical Fermi surface for $d_{xz}$ and $d_{yz}$ orbitals while two circular Fermi surfaces exist for the $d_{xy}$ orbital.

Addressing the above mentioned issues needs high quality data as well as a different approach. As for the experimental side, we take data in a different Brillouin zone (BZ) with all possible combinations of light polarizations to look for any missing bands. Meanwhile, our strategy is look for local orbital angular momentum (OAM) which could play a role in the surface band splitting. In that case, the existence of OAM in the surface states, which could be indirectly probed by circular dichroism (CD) ARPES experiment, may be used to explain the linear term in the surface band splitting within a Rashba-like model~\cite{S.R Park}. Our experimental results reveal additional subbands for $d_{xz/yz}$ and are suggestive of a chiral OAM structure, providing possible clues for the Rashba-like splitting effect.

\section{Methods}
Single crystals of lightly Nb doped (0.05 weight $\%$) SrTiO$_3$ (MTI, USA) were cut into $5\times 2\times 0.5$ mm$^{3}$  pieces and mounted on a custom designed sample holder with a scratch line for cleavage (Fig. 1(a),(b)). Samples were cleaved \emph{in situ} and measured at 18 K. Flat and shiny surfaces with areas $> 1\times 0.2$~mm$^{2}$ as required for the ARPES experiment were obtained (Fig. 1(c)). The data presented in this paper was taken at the beam line I05 of the Diamond Light Source in the United Kingdom. Preliminary experiments were performed at beamline 4A1 at Pohang Light Source, BL21B1 at National Synchrotron Radiation Research Center, and I3 at MAX IV Laboratory. 51 eV photons with linear and circular polarizations were used. The experimental chamber is equipped with a Scienta R4000 analyzer with a vertical slit. The base pressure in the measurement chamber was better than $1.2\times 10^{-10}$ mbar. The energy and angular resolutions were 15 meV and 0.25$^{\circ}$, respectively. The experimental geometry is schematically shown in Fig. 1(d). When we have xz-plane as the mirror plane and linear vertical (LV) polarized light (Fig. 1(d)), we expect to see only the $d_{xy}$ and $d_{yz}$ bands. For the linear horizontal (LH) case, we expect to see $d_{xz}$ band only. For a Fermi surface map, $\theta$ (Fig. 1(d))) changes as much as 16 degrees, which results in a slight change in the polarization. The effect of such polarization change is estimated to be about 2$\%$ which is not significant at the qualitative level of our discussion.
\begin{figure}[t]
\centering
\epsfxsize=8.0cm
\epsfbox{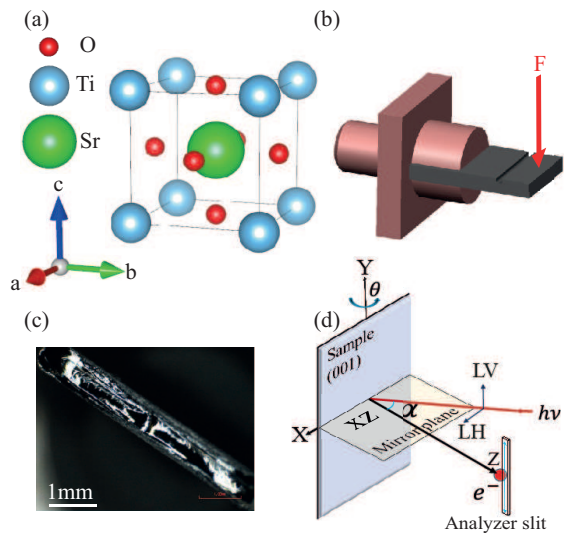}
\caption{(a) Crystal structure of SrTiO$_3$ at room temperature. (b) Schematic of the sample holder. Force F is exerted normal to the sample side-surface for cleaving. (c) Microscope image of a cleaved (001) surface ($0.5\times 5$ mm$^2$). (d) Schematic diagram for the experimental geometry; The analyzer slit is in the vertical direction; LV and LH show the case for vertical and horizontal polarization of the light with $\alpha=$50$^{\circ}$; Synchrotron beam (red line) with the energy $h$$\nu$ comes in in the xz-plane; Photoelectrons (shown by a red sphere) are collected by the analyzer with the slit in the vertical direction. For Fermi surface mapping, $\theta$ is rotated.} \label{figure1}
\end{figure}

\section{Subband structure of $d_{xz/yz}$ orbitals}

\begin{figure*}[t!]
\centering
\epsfxsize=15.0cm
\epsfbox{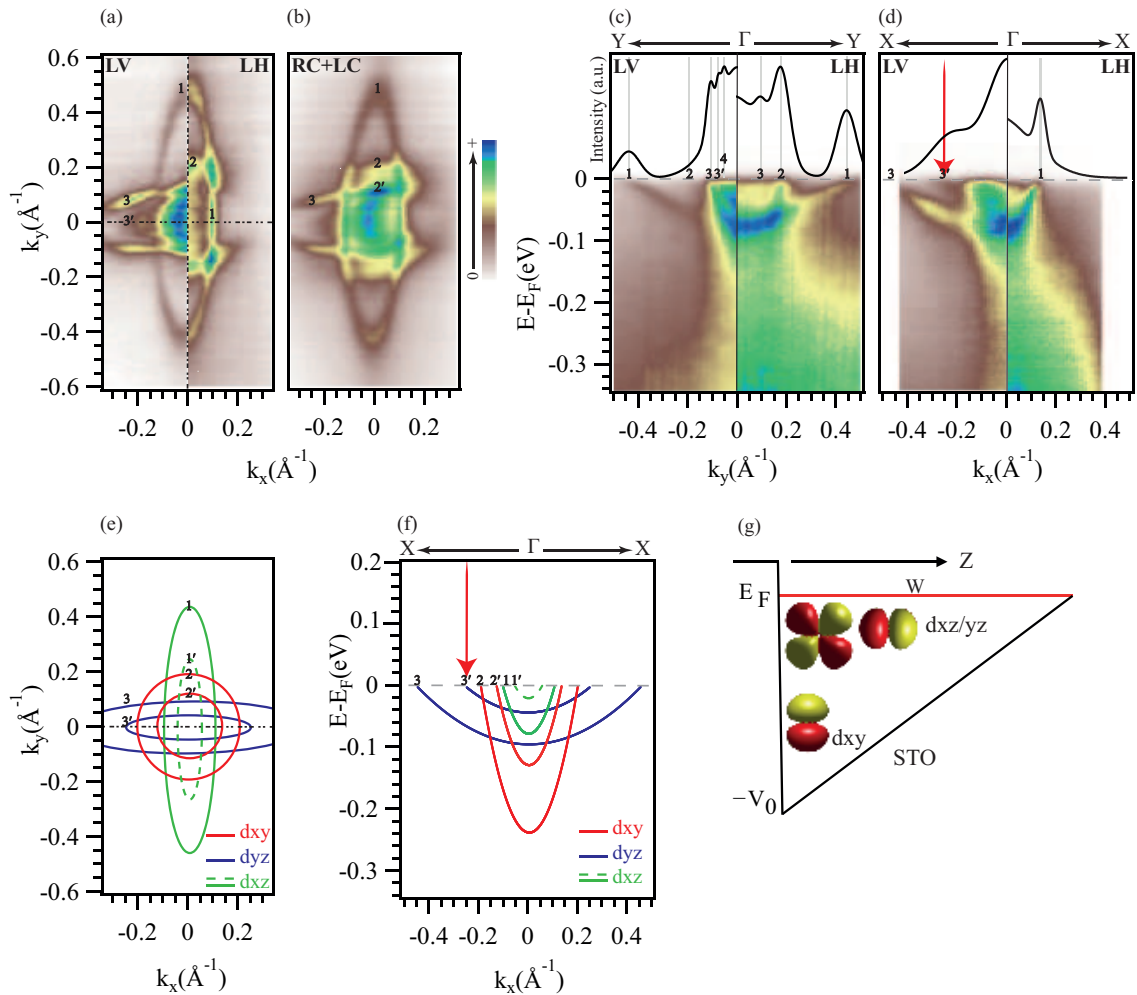}
\caption{ Fermi surface map of STO(001) surface taken with (a) LV and LH polarized light, and (b) sum of the data taken with RC and LC polarizations. (c) The cut along the $k_y$ direction taken with LV and LH polarizations. Fitted MDC at the Fermi level  case is plotted at the top with a black curve. The numbered grey vertical lines indicate peak positions in the MDCs. Primed numbers show inner subbands. (d) Data along the $k_x$ direction with two linear polarizations. The inner subband for $d_{yz}$ orbital is indicated by vertical red arrow. (e) Schematic diagrams for the Fermi surface extracted from the ARPES. The dashed green ellipsoid (1$'$) is expected based on the symmetry consideration but was not clearly observed in the data. (f) Schematic band dispersions along the $k_x$ direction. (g) Schematic of the confining potential on the (001) surface and illustration of the energy positions of the bands. In-plane $d_{xy}$ orbital are located at the bottom of the potential well while out-of-plane $d_{xz/yz}$ orbitals have a higher energy with the extended orbitals along the $z$ direction. } \label{figure2}
\end{figure*}

Figure 2 summarizes our ARPES data from STO (001) surface taken in the first BZ with all possible polarizations which are necessary to discern all the bands. We first discuss the Fermi surface map data in Fig. 2(a) and (b). Looking at the LV data along the $k_{x}$ direction in Fig. 2(a), we find at least two elliptical Fermi surfaces with semi-major axes of 0.45 and 0.25 $\mathrm{\AA}^{-1}$ and semi-minor axes of 0.11 and 0.07 $\mathrm{\AA}^{-1}$ (shown by number 3 and 3$'$). Considering the light polarization and shape, the two elliptical Fermi surfaces are from the $d_{yz}$ orbital and we will call them outer and inner subbands, respectively.  On the other hand, the data taken with LH shows that the two Fermi surfaces are suppressed along the $k_{x}$ direction while they appear along the $k_{y}$ direction. These observations reveal that the two elliptical Fermi surfaces are indeed from the $d_{yz}$ orbital. The data taken with circularly polarized light in Fig. 2(b) (sum of data taken with right and left circularly (RC and LC) polarized light) shows all the subbands even though they are not clearly discerned.

Figure 2(c) shows high-symmetry cuts along the $k_{y}$ direction (vertical dash-dot line in Fig. 2(a)) taken with LV (left) and LH (right) polarized light. Fitted momentum distribution curves (MDC) at the Fermi level are shown at the top for two polarizations. The numbered grey vertical lines mark the momentum positions where we have peaks in the MDC or Fermi level crossing. The LV data reveals a few heavy and light bands. The heavy band (number 1) is characterized as $d_{xz}$ with the band minimum located at 80 meV and Fermi wave vector ($k_F$) of 0.45 $\mathrm{\AA}^{-1}$. The suppressed intensity of this band is due to the polarization dependence of orbitals~\cite{Damascelli}. In addition, there are at least two light bands with their minima located at 45 and 100 meV with $k_F=$0.07 and 0.11 $\mathrm{\AA}^{-1}$ (number 3$'$ and 3), respectively. Considering the selection rules, we attribute these to $d_{yz}$ subbands. The peaks in the MDC curve suggest that there are two other bands with $k_F=$0.05 and 0.21 $\mathrm{\AA}^{-1}$. The first one belongs to $d_{yz}$ or $d_{xz}$ orbital, but from our polarization dependence data we can not say which one is and label it as number 4. The band with $k_F=$0.21 $\mathrm{\AA}^{-1}$ (shown by number 2) is the outer subband of $d_{xy}$ orbital. Right-hand side of Fig. 2(c) shows at least three Fermi level crossings. The heavy band from $d_{xz}$ (number 1), one light band from $d_{xy}$ orbital (number 2) and one located at $k_F=$0.11 $\mathrm{\AA}^{-1}$ (number 3) from $d_{yz}$ orbital.

Figure 2(d) shows data along the $k_x$ direction. The data taken with LV provides more evidence for the subband structure for $d_{yz}$ orbital. The LV data on the left hand side shows two heavy bands from $d_{yz}$. Their band minima are located at 45 and 100 meV with $k_F=$0.25 and 0.45 $\mathrm{\AA}^{-1}$ (number 3$'$ and 3), respectively. The band from the inner elliptical band (3$'$) is indicated by a red arrow. In the data taken with LH polarization depicted in Fig. 2(d), these two heavy bands are suppressed while the $d_{xz}$ intensity (1) is relatively strong as expected from the polarization dependence.

Figure 2(e) is the schematic Fermi surface determined based on our ARPES results. There are two elliptical Fermi surfaces for $d_{xz/yz}$. Note that the dashed green ellipsoid (1$'$) is missing in our ARPES data but we expect it based on the symmetry consideration. With the newly found Fermi surface topology, we estimate the carrier density from the area enclosed by the Fermi surfaces  to be $n_{2D}$=$A_F$/2$\pi^2$$\approx$$3\times$$10^{14}$ $cm^{-2}$. Shown in Fig. 2(f) is the schematic summary of the band structure. The subband structure for $d_{yz}$ orbital deduced from Fig. 2(a) and (d) is drawn with blue lines. The effective mass for the $d_{yz}$ bands in the $k_x$ direction is found to be $m^{\ast}\approx10m_e$ consistent with previous reports~\cite{Meevasana,Santander-Syro1,Plumb}. As mentioned above, a discernible inner $d_{xz}$ subband was not observed in the LH data in Figs. 2(a) and (d). We therefore plot the expected but missing inner band with dashed green parabola. The effective mass for the observed $d_{xz}$ band in the $k_{x}$ direction is estimated to be $m^{\ast}$$\approx$0.6$m_e$. $d_{xy}$ subbands obtained based on LH and RC+LC data are shown with red lines (number 2 and 2$'$) with the energy minima located at 130 and 240 meV.

We wish to emphasize two points here. First, the actual Fermi surface can be more complicated than the one illustrated in our simple schematic figure as we did not consider possible hybridization between different orbitals at the crossing points due to distorted crystal structure and spin-orbit coupling. Such effects cannot be resolved in the experimental data due to the limitated resolution. Second, there is a structural phase transition from cubic to tetragonal at 110 K,~\cite{Lytle,Unoki,Mattheiss2,Chang2,Santander-Syro1} and to orthorhombic below 65 K~\cite{Sakudo}. As a result, the degeneracy between $d_{xz}$ and $d_{yz}$ bands is lifted at the measurement temperature (18 K) and the effect appears as the non-degeneracy in the band dispersion at the $\Gamma$-point in Fig. 2(f).

Comparing our data with previously published data~\cite{Meevasana,Santander-Syro1,King}, one may ask why only a single band has been observed for $d_{xz/yz}$ before. Our data is collected in the first BZ where only the circular subbands from $d_{xy}$ orbital were observable. One possible explanation is that experimental conditions including photon energy, light polarization, and experimental geometry allowed us to observe the extra band. For example, the orientation of the $d_{xz/yz}$ orbitals inside surface potential well (Fig. 2(g))  means more three dimensional (3D) states and possibly a weak $k_z$ dispersion as it was suggested previously~\cite{Plumb}. Our photon energy dependent data indeed show a weak but discernible effect (not shown). In such a case, the selection of the photon energy could greatly affect the cross section. Another possibility is that the second $d_{xz/yz}$ band was not formed in other cases due to the finite depth of the potential well. However, the fact that the $d_{xy}$ band bottom is similar to the previously reported value suggests that the quantum well is similar to previous cases. This makes the latter explanation less likely.

\begin{figure*}[t!]
\centering
\epsfxsize=15.0cm
\epsfbox{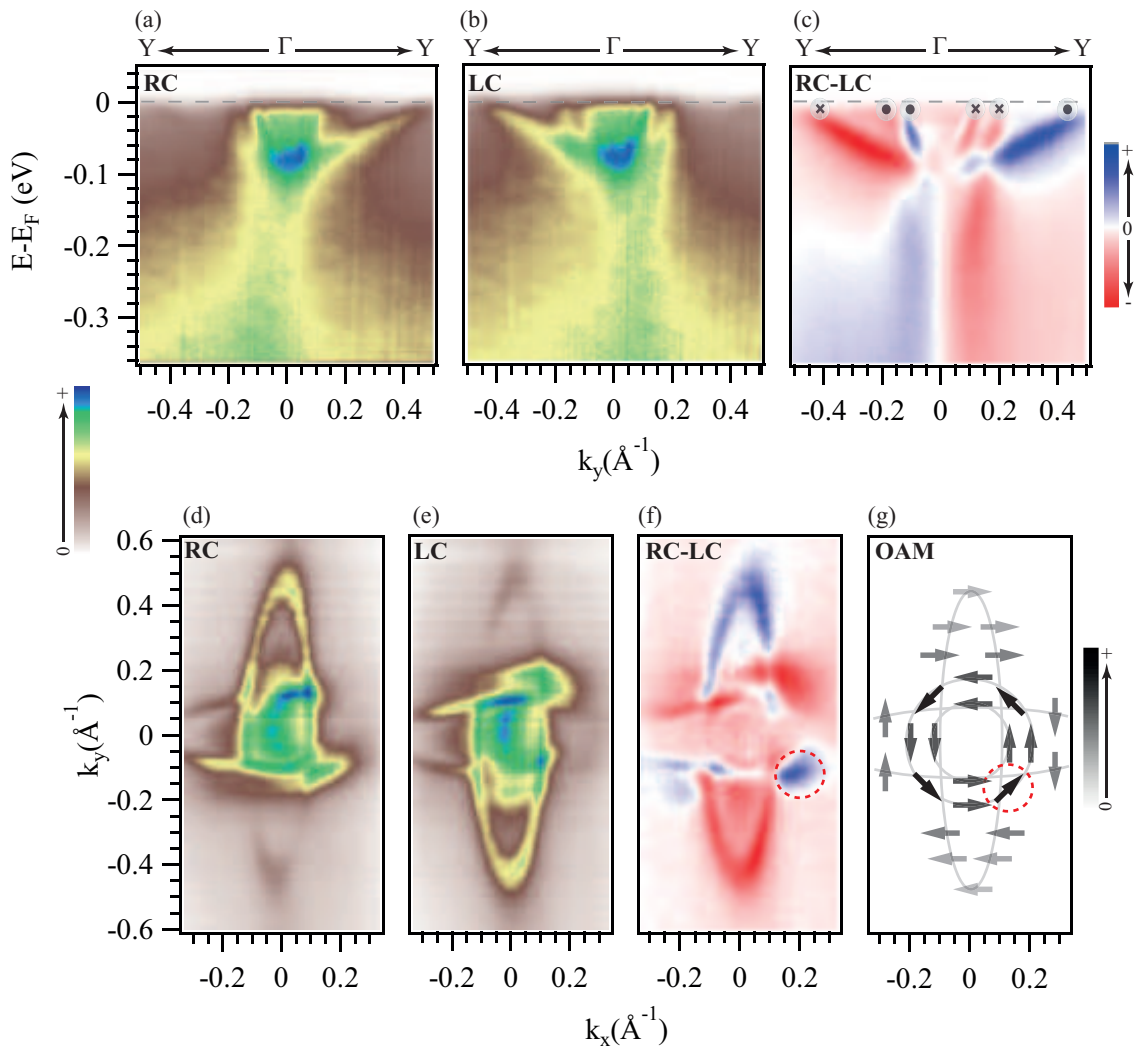}
\caption{Y-$\Gamma$-Y cut taken with (a) RC polarized light and (b) LC polarized light. Dashed gray lines represent the Fermi energy. (c) CD signal which is defined as the difference between RC and LC. A possible configuration of the OAM direction near Fermi energy is shown. Fermi surface maps of (d) RC polarized light, (e) LC polarized light, and (f) CD signal. Red dashed circle indicates the crossing points where multiorbital bands exist due to hybridization. (g) A possible configuration of the chiral OAM structure based on the CD signal.} \label{figure3}
\end{figure*}

\section{Chiral OAM structure}

With all subbands characterized in the previous section, we shift our focus to the study of the surface band splitting mechanism. As mentioned before, different approaches such as unconventional Rashba model were used to explain the surface band splitting. Unfortunately, the split bands and spin structure are not expected to be observed by ARPES due to the small Rashba parameter of STO~\cite{King}. On the other hand, it was argued that, in the presence of inversion symmetry breaking, the multi-orbital bands can lead to chiral OAM structures.~\cite{Panjin Kim} In that approach, a net OAM, which is defined as the sum of two OAM vectors of inner and outer Rashba split bands, is expected. Noting the suggestion that useful information on the OAM can be obtained from CD-ARPES signal,~\cite{J.H. Park} CD-ARPES on the STO surface state may shed light on the issue.

Before discussing the usefulness of the CD-ARPES technique, we discuss the origin of the CD-ARPES signal as there are different views on the origin of the CD signal~\cite{S.R Park,Dubs,Westphal,Ishida}. CD signal refers to the difference in the reaction to right and left circularly polarized light. The difference may come from experimental geometry~\cite{Dubs,Westphal,Ishida} and OAM in the initial~\cite{S.R Park} and/or final states~\cite{Scholz,Vidal}. To estimate the contribution from experimental geometry, we took CD-ARPES data on a polycrystalline gold sample with the same experimental condition. As polycrystalline gold is expected to have no OAM, any observed signal would be from experimental geometry. The normalized CD (NCD) from gold sample, defined as NCD=(RC-LC)/(RC+LC), shows a weak signal less than 5$\%$ which is much smaller than the typical NCD value from STO of about 60$\%$. In addition, the complicated CD pattern in Figs. 3(c) and (f) also shows that it is not from the geometrical effect for which a simple asymmetric CD pattern is expected. On the other hand, with the relatively high photon energy used in the experiment, the final state is expected to be close to a free electron-like state. Therefore, we assume that the major part of the CD pattern is determined by the OAM of initial states, and discuss the CD-ARPES within this interpretation.

We plot our CD-ARPES results in Fig. 3. Figures 3(a) and (b) show high symmetry cuts along the $\Gamma$-Y direction with RC and LC polarized light, respectively. It is clear that the RC data has higher intensity on the right while the LC data shows an opposite behavior. The CD signal defined as CD=RC-LC is shown in Fig. 3(c). The sign change in the CD signal may be attributed to the reversal of the OAM direction. An OAM pattern consistent with the CD data is marked in the figure. We point out that the pattern with OAM direction reversed is also consistent with the CD data. Fermi surface maps for CD-ARPES study are depicted in Figs. 3(d) - (f).  Due to the complicated band structure, a clear identification of the bands can be made only for two large elliptical $d_{xz/yz}$ and two circular $d_{xy}$ Fermi surfaces in the CD Fermi surface map in Fig. 3(f).

Figure 3(g) shows an OAM pattern on the Fermi surface that is consistent with the CD-ARPES data. To deduce the OAM pattern, we use the following simple rules. First, we take the positive (blue color) and negative (red color) signal in Fig. 3(f) as rightward and leftward OAM vectors, respectively. Second, any OAM pattern should preserve the fourfold symmetry of the Fermi surface.  As for the magnitude of the OAM vectors, we note that at the crossing points where multiorbital bands exist the magnitude of the OAM is larger due to the enhanced spin orbit coupling effect.~\cite{King,Panjin Kim} Finally, near the $k_{y}$=0 line which coincides with the mirror plane, CD is expected to be very weak. In that case, we have applied the symmetry rule to deduce OAM vectors. Here, we point out that the OAM texture we have shown in Fig. 3(g) is not unique. For example, the pattern with reversed OAM vectors is still valid. Therefore, the OAM texture in Fig. 3(g) is not a unique solution but a consistent one. However, it is sufficient for the discussion to follow.

Our observation suggest that in-plane OAM vectors for $d_{xz/yz}$ and $d_{xy}$ orbitals have opposite chiralities while the chiralities for the two $d_{xy}$ subbands are the same. The non-zero OAM implies that the OAM of the two Rashba bands do not cancel each other. This matches theoretical reports~\cite{King,Panjin Kim} that predict the same OAM directions for the Rashba bands near crossing points (shown with dashed red circles in Fig. 3(f) and (g)) due to orbital mixing. Far from crossing points, Ref.~\cite{King} predicts OAM vectors of the $d_{xz/yz}$ Rashba bands to have opposite directions while Ref.~\cite{Panjin Kim} suggests for the same OAM directions for the $d_{xz/yz}$ Rashba bands. In this sense, our non-zero CD-signal for $d_{xz/yz}$ bands is in a better agreement with predictions of Ref.~\cite{Panjin Kim}. On the other hand, the $k$ cubic term due to multiorbital effects~\cite{Zhong} is much smaller than the linear term~\cite{Nakamura,Panjin Kim} and thus cannot be detected in our CD-ARPES. Within the interpretation discussed so far, we should see more pronounced signal near crossing points and weaker CD signal away from these points as seen in Fig. 3(f). The observation of OAM also suggests that the Rashba-related band splitting can be understood within the so-called orbital Rashba effect~\cite{Panjin Kim,J.H. Park,S.R Park} rather than the conventional Rashba model~\cite{Rashba,Bihlmayer}.

Our results could shed light on the recent spin-ARPES measurements on the surface states of STO.~\cite{Santander-Syro2,Walker3} Based on their experimental observations, authors of Ref.~\cite{Santander-Syro2} suggested that a giant non-Rashba type splitting of 100 meV with opposite spin chiralities in the $d_{xy}$ subbands exists. On the other hand, a more recent spin-ARPES measurement~\cite{Walker3} shows no such giant spin splitting. While CD-ARPES is not a spin sensitive technique and these inconsistent results should be further resolved experimentally and theoretically, we believe our result may have  some implication on the issue. First of all, our CD-ARPES data, being consistent with the predictions of Ref.~\cite{Panjin Kim}, suggest the same OAM direction for the two Rashba bands, which in turn tells us that the spins in the two Rashba bands are pointing oppositely. Therefore, the net spin from a $d_{xy}$ subband should be zero, consistent with the result reported in Ref.~\cite{Walker3}. Why are there two conflicting results then? A possible solution comes from the earlier suggestion that, in systems with chiral OAM texture, the observed spin polarization of surface states could be strongly affected by the light polarization or experimental geometry~\cite{Cheol-Hwan Park,Jozwiak,Sanchez-Barriga}. While the net spin is zero, the measured spin polarization can be non-zero if the two spin signals from the two bands are affected differently by the polarization of the light or experimental geometry. However, we should point out that if the observed non-zero spin comes from the above mentioned effect, we expect similar effects for the two $d_{xy}$ subbnads. Therefore, within our interpretation, it is still hard to explain the fact that the two $d_{xy}$ subbands show opposite chiralities while our data show the same OAM chiralities.


\section{Conclusion}

In summary, we have performed ARPES measurements on the 2DEG on the surface of STO single crystal using linearly and circularly polarized light. Our measurements in the first BZ using LV polarized light reveals subbands for out of plane $d_{xz}$ orbital with $k_F=$0.25 and 0.45 $\mathrm{\AA}^{-1}$, which implies $d_{yz}$ orbital also has a subband considering the four fold symmetry. Therefore, subbands exist not only for $d_{xy}$ but also for $d_{yz/zx}$ orbitals. In addition, CD results suggest for a chiral OAM texture in momentum space. It supports the theoretical predictions~\cite{Panjin Kim} of same OAM directions for the two Rashba bands near crossing points where multiorbital bands exist, which implies a better consistency with an orbital Rashba~\cite{Panjin Kim,J.H. Park,S.R Park} than a conventional Rashba model~\cite{Ishida,Rashba}.

\section*{Acknowledgment}

We are grateful for helpful discussions with Jung Hoon Han and Young Jun Chang. We thank Wonshik Kyung for his role in preliminary experiments. The work was supported by  IBS-R009-G2. S.S., S.C., H.Y. and G. H. acknowledge were supported by Yonsei university, BK21 program. The data presented in this paper was taken at beamline I05, Diamond Light Source under proposal SI11445.


\end{document}